\documentclass[11pt]{article}

\usepackage[margin=1in]{geometry}
\usepackage{times}
\usepackage{microtype}
\usepackage{amsmath}
\usepackage{amssymb}
\usepackage{booktabs}
\usepackage{multirow}
\usepackage{graphicx}
\usepackage{float}
\usepackage{subcaption}
\usepackage{url}
\usepackage{hyperref}
\usepackage[numbers,sort&compress]{natbib}

\hypersetup{
  colorlinks=true,
  linkcolor=blue,
  citecolor=blue,
  urlcolor=blue
}

\title{Can Graph-Based Microservice Performance Detection Be Used for Microservice Intrusion Detection?}

\author{
Yunjian Ma\\
Independent Researcher\\
\texttt{ }
}

\date{}

\begin{document}

\maketitle

\begin{abstract}
Microservice systems expose rich telemetry streams, including metrics, logs, and distributed traces. Existing performance anomaly detection methods increasingly model these systems as graphs, where nodes represent services and edges represent runtime dependencies. This paper asks whether graph-based microservice performance detection can also serve as a foundation for microservice intrusion detection. We deploy a Docker Compose based synthetic e-commerce microservice benchmark, run 50 controlled trials across five attack types under normal workloads, and collect metrics, logs, and distributed traces. Each request trace is converted into a request-level invocation graph with multi-modal node features derived from timestamped logs and per-service performance metrics. As a first baseline, we train a two-layer graph convolutional network (GCN) for 6-way classification (normal plus five attack types) over 21,438 request graphs. The GCN achieves 96.2\% test accuracy with a macro F1 of 0.955 under a graph-level random split. We then conduct modality ablation, trial-level split evaluation, non-graph baseline comparison, runtime analysis, t-SNE visualization, confusion-matrix analysis, and error-case inspection. The stricter trial-level results show that trace structure alone is insufficient, logs and metrics improve detection, and strong flattened baselines currently outperform the shallow GCN on the engineered feature set.
\end{abstract}


\section{Introduction}

Microservice architectures decompose applications into independently deployed services that communicate through HTTP, gRPC, message queues, databases, and service meshes. This decomposition improves modularity and deployment velocity, but it also makes operational and security monitoring more difficult. A single user request may traverse many services, and a local failure or compromise can propagate through the dependency graph.

Performance anomaly detection and intrusion detection are often treated as separate problems. Performance monitoring focuses on latency, throughput, resource consumption, and error rates, while intrusion detection focuses on malicious behavior, unauthorized access, suspicious payloads, and adversarial movement. In microservice systems, however, these concerns can overlap. Some attacks create visible performance symptoms: denial-of-service attacks increase request rates and latency; brute-force login attempts alter authentication failure patterns; data exfiltration may change read volume or response size; lateral movement may introduce unusual service-to-service paths.

This paper investigates the following question:

\begin{quote}
Can graph-based microservice performance detection be used for microservice intrusion detection?
\end{quote}

The goal is not to claim that performance anomaly detection is a complete intrusion detection system. Instead, this paper studies when performance-oriented graph methods are useful for security detection, when they fail, and how logs and traces can enrich the graph representation.

The contributions are:

\begin{itemize}
  \item A reproducible experimental design for evaluating microservice intrusion scenarios in an observable environment, with 50 automated trials across five attack types.
  \item A request-level invocation graph representation that maps one distributed trace, related logs, and nearby service metrics into a single graph sample with 25-dimensional node features.
  \item A baseline GCN classifier achieving 96.2\% test accuracy across six classes under a graph-level random split, establishing an optimistic within-distribution reference point.
  \item An exploratory evaluation comparing modality ablations, trial-level split generalization, non-graph baselines, runtime costs, t-SNE visualization, and confusion-matrix error patterns.
\end{itemize}

\section{Background and Research Questions}

Modern microservice platforms expose three telemetry pillars: metrics, logs, and traces. Metrics describe resource use and latency, logs contain discrete application and security events, and distributed traces reveal request paths across service boundaries. Performance anomaly detection and intrusion detection are often studied separately, but in microservice systems they overlap: an attack may change request rate, latency, error propagation, authentication failures, or the service dependency path. This motivates a request-level graph representation that preserves service dependencies while incorporating operational and security-relevant signals.

This paper is organized around four research questions:

\begin{description}
  \item[RQ1:] Can a graph neural network classify a single microservice request invocation graph as normal or intrusive?
  \item[RQ2:] Which intrusion categories are visible through request call-path, service performance, and log signals?
  \item[RQ3:] Does adding timestamped logs and node-level metrics improve detection beyond trace-only invocation graphs?
  \item[RQ4:] How do graph-based methods compare with non-graph baselines on detection quality and runtime cost?
\end{description}

\section{Related Work}

\textbf{Microservice anomaly detection and troubleshooting.} Distributed tracing systems such as Dapper made request-level execution paths practical to collect in large distributed systems \citep{sigelman2010dapper}. Building on such telemetry, MicroRank localizes end-to-end latency issues using trace-based spectrum analysis \citep{yu2021microrank}, while ServiceRank combines anomaly detection and root-cause identification for large-scale microservice architectures \citep{ma2022servicerank}. Eadro is especially close to this work: it models traces, logs, and KPIs jointly for end-to-end anomaly detection and root-cause localization in microservices \citep{lee2023eadro}. Our work follows Eadro's multi-source observability motivation, but asks a narrower security-oriented question: when can graph-based performance detection serve as a useful foundation for intrusion detection?

\textbf{Graph-based microservice detection.} Several recent systems model traces and monitoring signals as graphs. DeepTraLog combines traces and logs into trace-event graphs and trains a graph-based deep learning detector \citep{zhang2022deeptralog}. TraceGra uses graph deep learning to model traces and performance metrics for unsupervised anomaly detection \citep{chen2023tracegra}. MSTGAD uses twin graphs and attentive multi-modal learning for microservice anomaly detection \citep{huang2023mstgad}. These works support the premise that graph structure can preserve service dependencies that flat feature vectors may lose. Our experiments add a trial-level split and non-graph baselines to test whether a simple request-level GCN actually outperforms flattened logs+metrics features in the current security benchmark.

\textbf{Graph neural networks and intrusion/anomaly detection.} The GCN used in this paper follows the graph convolutional formulation of \citet{kipf2017semi}; GraphSAGE and graph attention networks provide common alternatives for inductive and attention-based graph representation learning \citep{hamilton2017inductive,velickovic2018graph}. More broadly, graph anomaly detection surveys emphasize that graph structure is useful when abnormal behavior manifests as unusual relations or neighborhoods \citep{akoglu2015graph}. For security-oriented anomaly detection, DeepLog learns log event sequences for system anomaly detection \citep{du2017deeplog}, while Kitsune demonstrates online network intrusion detection with autoencoder ensembles \citep{mirsky2018kitsune}. These methods motivate our comparison between graph-based request representations, log/metric modalities, and simpler non-graph baselines.

\section{Methodology}

The experimental environment is a synthetic e-commerce microservice benchmark with six services: \texttt{frontend}, \texttt{auth}, \texttt{catalog}, \texttt{cart}, \texttt{order}, and \texttt{payment}. The attacker interacts with externally reachable application APIs; the study does not assume host compromise, kernel-level compromise, or control of the container runtime. We evaluate five implemented application-layer attack scenarios: HTTP flood, brute-force login, SQL injection probe, SSRF-like probe, and exfiltration simulation. Normal traffic exercises common user behaviors such as browsing, search, authentication, cart access, and order placement. Implementation and reproduction details are provided in the appendix.

The detection unit is a single request trace. For each request trace $r_i$, the experiments construct one service-level graph:

\begin{equation}
G_i = (V_i, E_i, X_i, y_i),
\end{equation}

where $V_i$ is the set of services involved in the request, $E_i$ is the service-level call relation observed in the trace and symmetrized for GCN message passing, $X_i$ is the node feature matrix, and $y_i$ is the request label. Edge attributes are not used in the reported models. The reported 25-dimensional node feature vector concatenates 16 log TF-IDF dimensions and 9 service metric dimensions. Trace information contributes the graph topology itself; the exploratory trace-only variant uses service-presence and coarse graph-shape features rather than span-duration or status-code attributes.

The graph model is a supervised two-layer GCN request classifier. Given graph $G_i$, the encoder computes node embeddings using graph convolution layers and applies mean pooling to obtain a graph-level request embedding. A small MLP maps the request embedding to class logits. For non-graph comparison, we train Random Forest and MLP classifiers on flattened logs+metrics request features. The reported metrics are accuracy, precision, recall, F1-score, macro F1, weighted F1, confusion matrices, t-SNE visualization, training time, and prediction latency.

\section{Results}

The dataset contains 50 trials and 21,438 labeled request-level invocation graphs across six classes. Each trial contains a normal window and an attack window; labels are assigned by trace-window overlap. Table~\ref{tab:dataset} summarizes the class distribution.

The baseline model is a two-layer GCN. Given a graph with adjacency matrix $A$ and degree matrix $D$, the normalized symmetric adjacency is $\hat{A} = D^{-1/2}(A + I)D^{-1/2}$. Node embeddings are computed as:

\begin{equation}
H^{(1)} = \text{ReLU}(\hat{A} X W^{(1)}), \quad
H^{(2)} = \text{ReLU}(\hat{A} H^{(1)} W^{(2)})
\end{equation}

A mean-pooling readout produces a graph-level embedding $h_G = \frac{1}{|V|}\sum_{v \in V} H^{(2)}_v$, which is passed through a 2-layer MLP classifier to produce class logits. The full hyperparameter and reproduction details are given in the appendix.

\begin{table}[H]
\centering
\caption{Dataset composition for the reported experiments.}
\label{tab:dataset}
\begin{tabular}{lrr}
\toprule
Class & Trials & Graphs \\
\midrule
normal & --- & 7,905 \\
http\_flood & 10 & 8,734 \\
bruteforce\_login & 10 & 1,563 \\
sql\_injection\_probe & 10 & 1,346 \\
exfiltration\_sim & 10 & 1,051 \\
ssrf\_probe & 10 & 839 \\
\midrule
Total & 50 & 21,438 \\
\bottomrule
\end{tabular}
\end{table}

\begin{table}[H]
\centering
\caption{Baseline GCN classification report (test set, $n=6,432$ graphs).}
\label{tab:baseline_results}
\begin{tabular}{lrrrr}
\toprule
Class & Precision & Recall & F1 & Support \\
\midrule
normal & 0.953 & 0.947 & 0.950 & 2,372 \\
http\_flood & 0.965 & \textbf{0.996} & 0.980 & 2,620 \\
sql\_injection\_probe & 0.957 & 0.983 & 0.969 & 404 \\
exfiltration\_sim & 0.968 & 0.946 & 0.957 & 315 \\
ssrf\_probe & \textbf{0.987} & 0.893 & 0.938 & 252 \\
bruteforce\_login & 0.990 & 0.885 & 0.935 & 469 \\
\midrule
Accuracy & \multicolumn{4}{c}{0.962} \\
Macro avg & 0.970 & 0.942 & 0.955 & 6,432 \\
Weighted avg & 0.963 & 0.962 & 0.962 & 6,432 \\
\bottomrule
\end{tabular}
\end{table}

Table~\ref{tab:baseline_results} reports the per-class and aggregate performance of the baseline GCN. The classifier achieves 96.2\% test accuracy with a weighted F1 of 0.962 and macro F1 of 0.955 under a graph-level random split. Because this split can include request graphs from the same trial in both train and test sets, this result is best interpreted as an optimistic within-distribution reference. The stricter trial-level split is reported in Table~\ref{tab:split_baselines}.

\textbf{Baseline observations.} The baseline reveals several patterns that inform subsequent experiments:

\begin{enumerate}
  \item \textbf{No overfitting:} train accuracy (95.8\%) is slightly below test accuracy (96.2\%), suggesting the model is under-parameterized relative to the data and could benefit from larger hidden dimensions.
  \item \textbf{Recall gap on minority classes:} \texttt{ssrf\_probe} (recall 0.893, 3.9\% of data) and \texttt{bruteforce\_login} (recall 0.885, 7.3\% of data) show a $\sim$10 percentage-point gap between precision and recall. The model is conservative on these classes---when it flags them it is almost always correct, but it misses roughly one in nine attacks. This is characteristic of class imbalance, as \texttt{normal} and \texttt{http\_flood} together account for 77.6\% of all graphs.
  \item \textbf{Performance-visible attacks are easiest:} \texttt{http\_flood} achieves near-perfect recall (0.996), consistent with its strong performance footprint.
  \item \textbf{Log-dependent attacks are detectable:} Both \texttt{sql\_injection\_probe} (F1 0.969) and \texttt{exfiltration\_sim} (F1 0.957) are well detected despite having weaker performance signatures, indicating that the TF-IDF log features capture security-relevant signals.
\end{enumerate}

These results establish the baseline. We next add modality ablation, trial-level splitting, non-graph baselines, and error analysis. Because these extended experiments compare many conditions quickly, the GCN runs in this section use 5 epochs rather than the 100-epoch baseline above. The absolute GCN scores should therefore be read as exploratory, while the relative modality and split patterns are still informative.

The original baseline uses a stratified graph-level random split. This tests within-distribution classification, but it may place request graphs from the same trial in both training and test sets. To test generalization across experimental runs, we add a trial-level split: all request graphs from a run id are assigned entirely to either train or test. This yields 15,035 training graphs and 6,403 test graphs.

\begin{table}[H]
\centering
\caption{Random versus trial-level split results for the extended experiment.}
\label{tab:split_baselines}
\begin{tabular}{lllrrrr}
\toprule
Split & Model & Input & Accuracy & Macro F1 & Train s & Pred ms/graph \\
\midrule
Random & GCN & logs+metrics & 0.903 & 0.911 & 15.38 & 0.0507 \\
Random & Random Forest & logs+metrics & 0.978 & 0.983 & 0.26 & 0.0019 \\
Random & MLP & logs+metrics & 0.973 & 0.976 & 0.52 & 0.0001 \\
Trial-level & GCN & logs+metrics & 0.879 & 0.777 & 15.61 & 0.0524 \\
Trial-level & Random Forest & logs+metrics & 0.971 & 0.980 & 0.21 & 0.0025 \\
Trial-level & MLP & logs+metrics & 0.964 & 0.974 & 0.46 & 0.0001 \\
\bottomrule
\end{tabular}
\end{table}

Table~\ref{tab:split_baselines} compares random and trial-level results. The 5-epoch GCN drops from 0.911 macro F1 under random split to 0.777 under trial-level split, showing that trial-level evaluation is substantially stricter. In contrast, flattened non-graph baselines remain strong under trial-level split, suggesting that the current engineered log and metric features are highly discriminative even without graph message passing.

\begin{table}[H]
\centering
\caption{GCN modality ablation under trial-level split.}
\label{tab:modality_ablation}
\begin{tabular}{lrrrr}
\toprule
Input modality & Accuracy & Macro F1 & Weighted F1 & Train s \\
\midrule
Trace structure & 0.660 & 0.405 & 0.621 & 15.57 \\
Logs only & 0.813 & 0.734 & 0.779 & 15.69 \\
Metrics only & 0.866 & 0.825 & 0.867 & 15.57 \\
Logs+metrics & 0.879 & 0.777 & 0.849 & 15.61 \\
\bottomrule
\end{tabular}
\end{table}

\begin{figure}[H]
\centering
\includegraphics[width=0.68\linewidth]{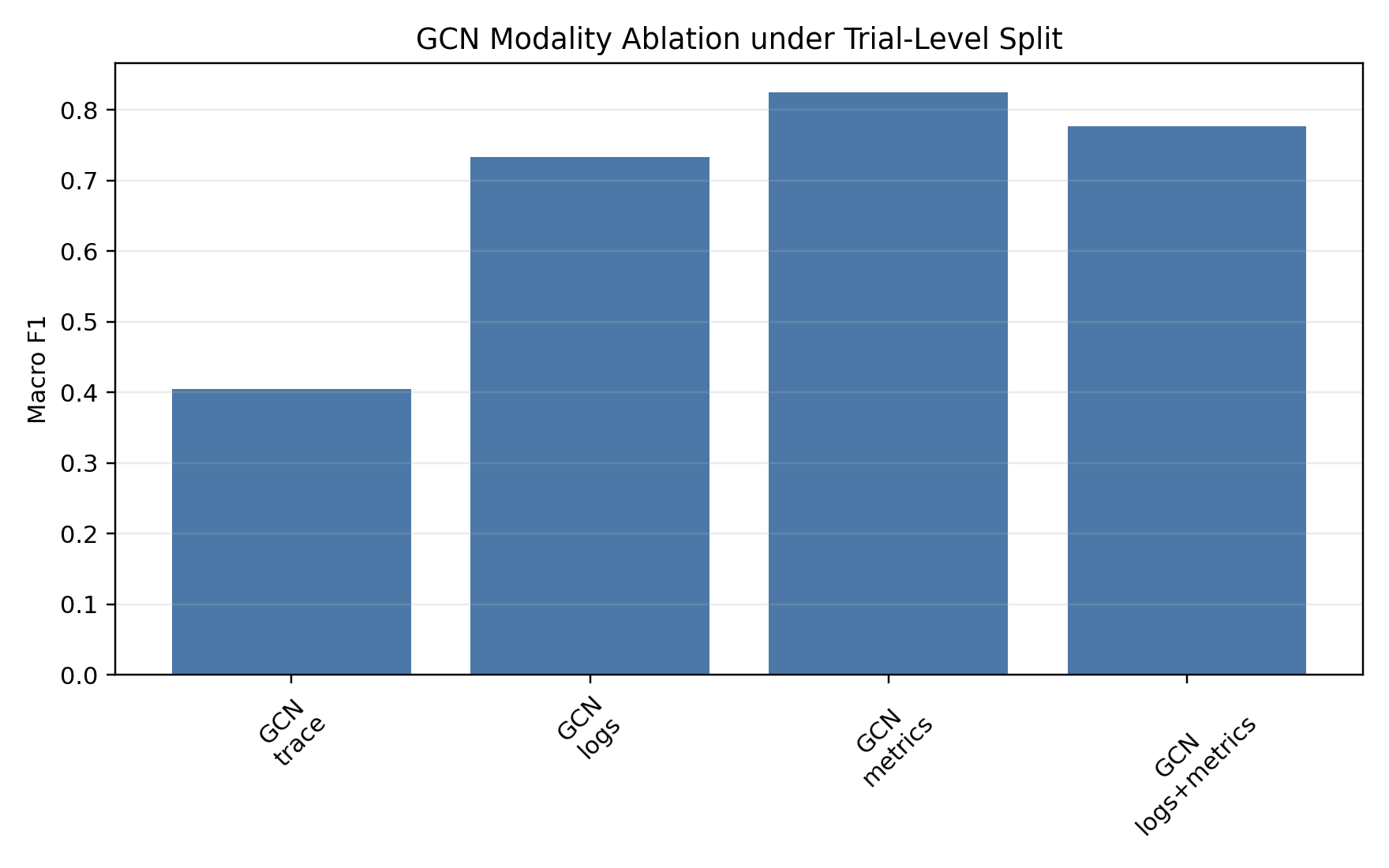}
\caption{Macro F1 for GCN modality ablation under trial-level split.}
\label{fig:modality_ablation}
\vspace{0.6em}
\includegraphics[width=0.62\linewidth]{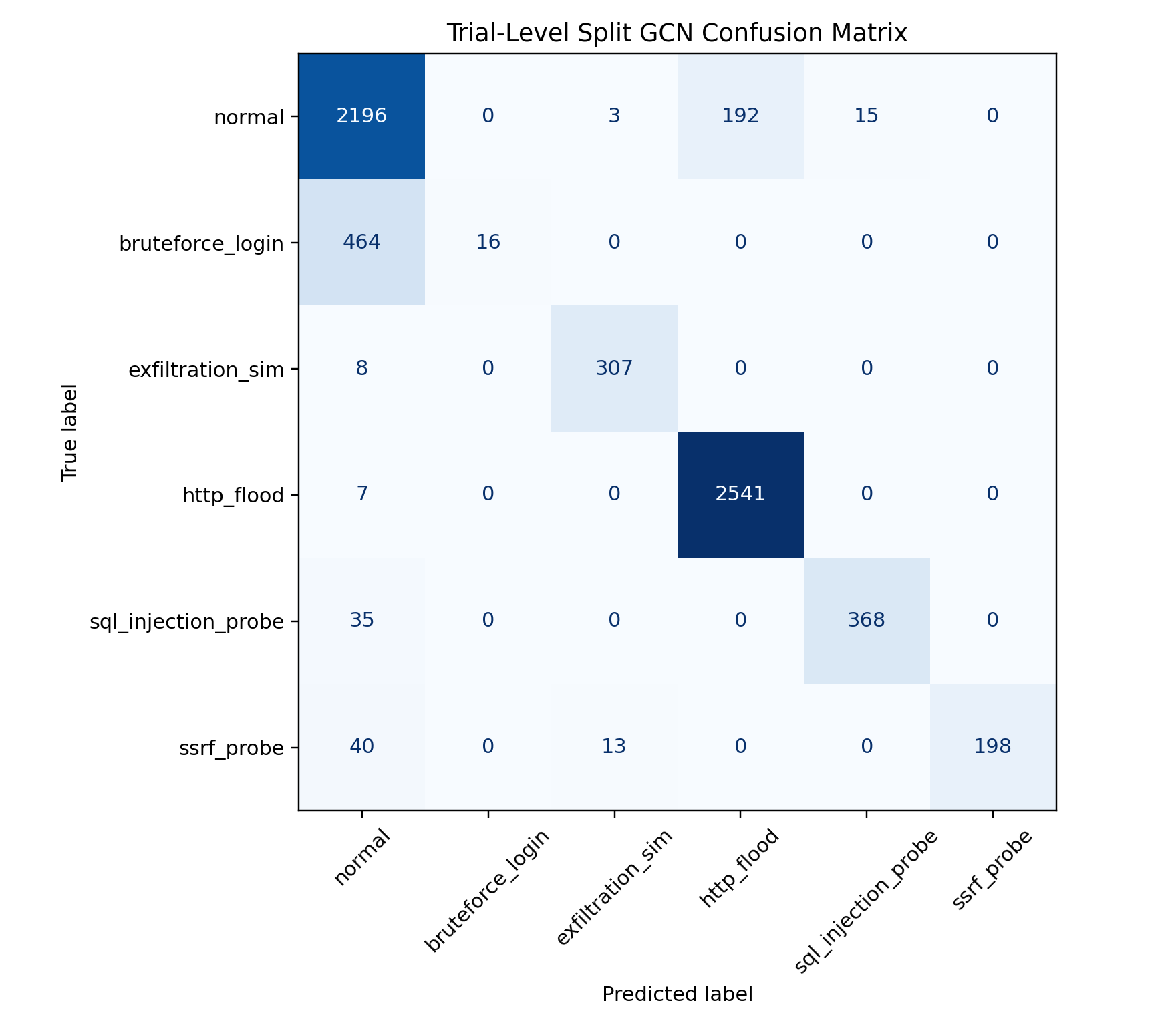}
\caption{Confusion matrix for trial-level GCN with logs+metrics.}
\label{fig:confusion_matrix}
\end{figure}

Table~\ref{tab:modality_ablation} reports GCN modality ablation under the trial-level split. Trace structure alone is insufficient (macro F1 0.405), which means call-path topology by itself does not identify all attack semantics in this benchmark. Logs improve macro F1 to 0.734, while metrics reach 0.825. The combined logs+metrics setting achieves the highest accuracy but lower macro F1 than metrics alone because the short 5-epoch GCN misses many brute-force requests. Figure~\ref{fig:modality_ablation} visualizes the same comparison as a bar chart, making the relative contribution of each telemetry view easier to inspect: trace-only features form the lowest bar, logs and metrics provide clear gains, and the all-feature setting improves accuracy but not macro F1. This supports a nuanced answer to RQ3: additional telemetry helps, but simply concatenating modalities is not guaranteed to improve every attack class without adequate training and class-balancing.

Figure~\ref{fig:confusion_matrix} shows the trial-level confusion matrix for the 5-epoch logs+metrics GCN. The dominant error pattern is brute-force login being predicted as normal. The sampled error records show repeated \texttt{auth|frontend} request graphs, usually with two nodes and one edge. This indicates that these attacks preserve a normal-looking request path and require stronger temporal or security-event features than a shallow request-level GCN currently uses.

Per-class recall reinforces this interpretation: HTTP flood remains easy (0.997 recall), exfiltration simulation is also strong (0.975), SQL injection probe remains detectable (0.913), and SSRF is moderate (0.789). Brute-force login recall falls to 0.033 in the short exploratory GCN run. The main lesson is that request-level graphs capture performance-visible and log-visible attacks, but repeated authentication failures need either class weighting, sequence aggregation across requests, or explicit failed-login counters.

\begin{figure}[H]
\centering
\includegraphics[width=0.78\linewidth]{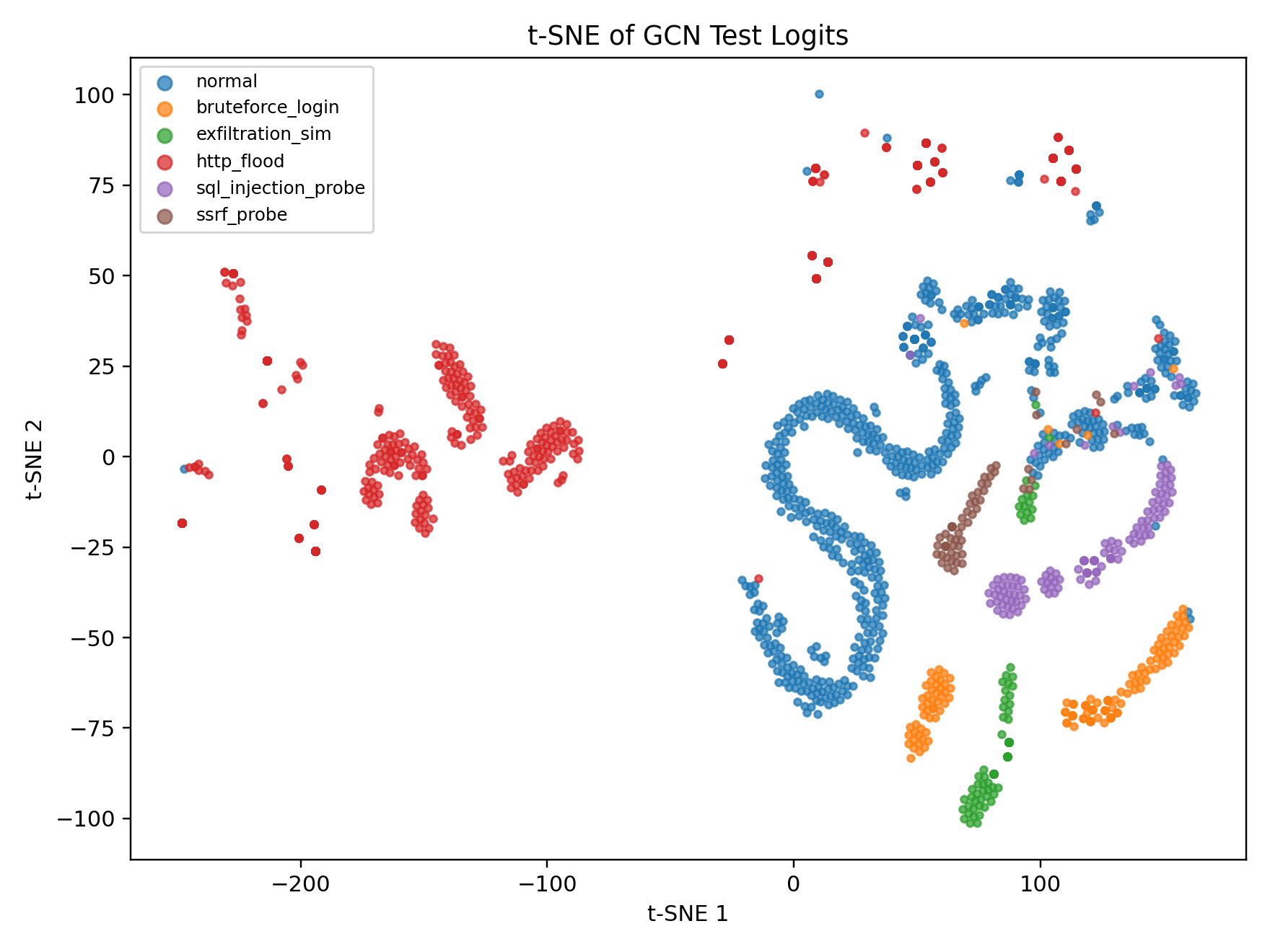}
\caption{t-SNE projection of trial-level GCN test logits. Separation is strongest for high-signal classes and weaker near normal/brute-force boundaries.}
\label{fig:tsne}
\end{figure}

\begin{figure}[H]
\centering
\begin{subfigure}[t]{0.48\linewidth}
\centering
\includegraphics[width=\linewidth]{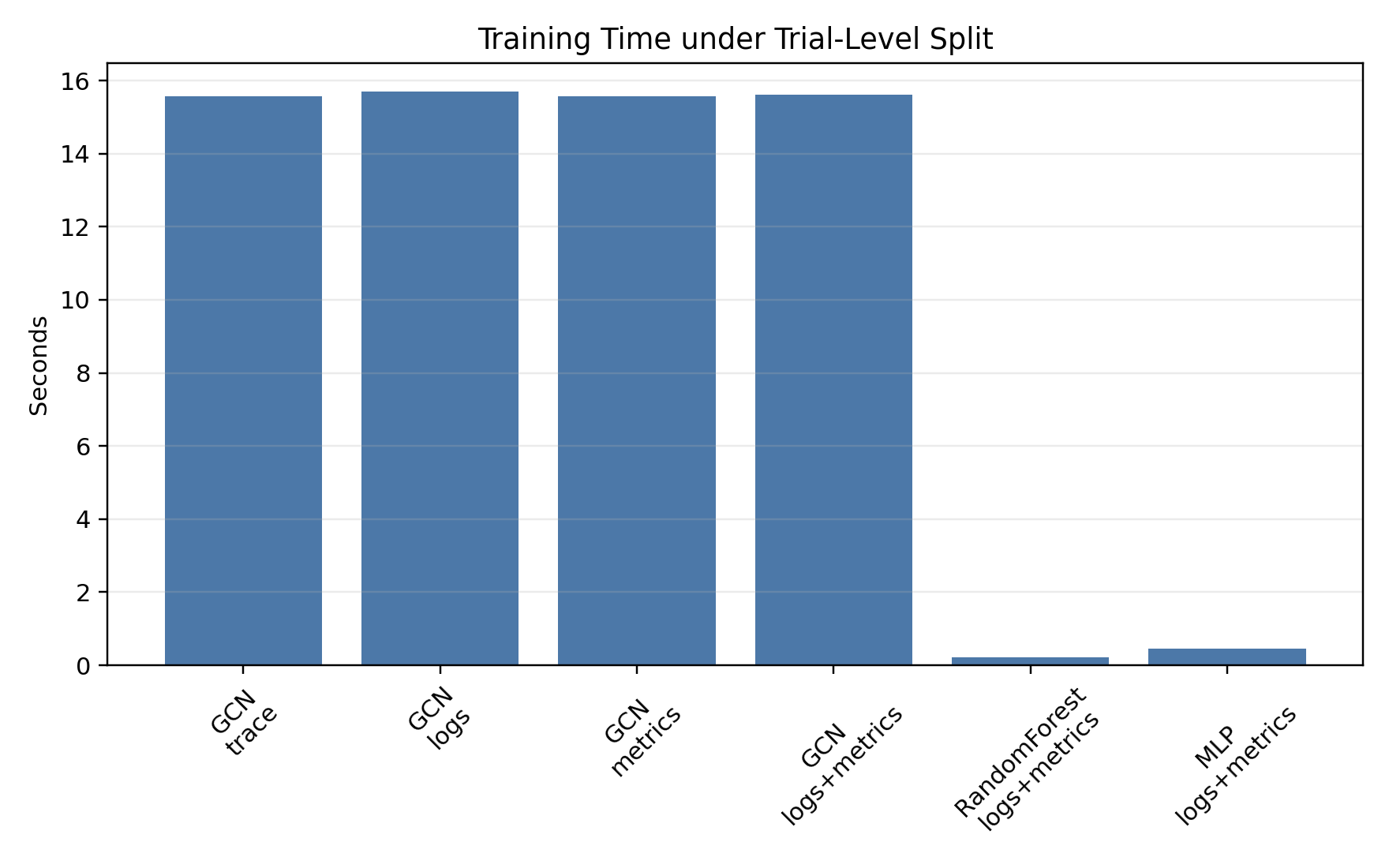}
\caption{Training time.}
\label{fig:train_speed}
\end{subfigure}\hfill
\begin{subfigure}[t]{0.48\linewidth}
\centering
\includegraphics[width=\linewidth]{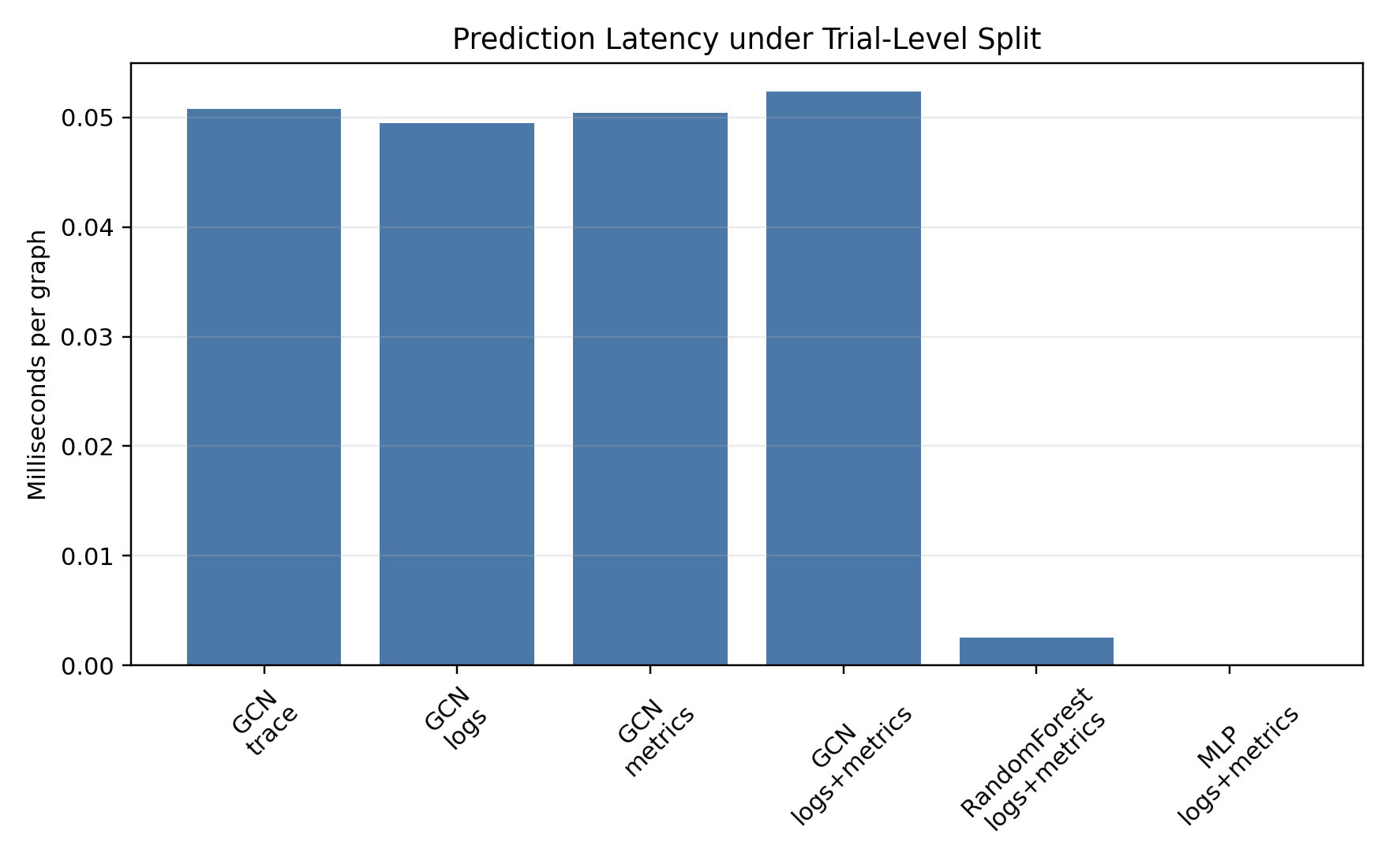}
\caption{Prediction latency.}
\label{fig:predict_speed}
\end{subfigure}
\caption{Runtime cost comparison under trial-level split.}
\label{fig:runtime_cost}
\end{figure}

Figure~\ref{fig:runtime_cost} compares training and prediction cost. The simple GCN trains in roughly 15.6 seconds per modality for 5 epochs and predicts in about 0.05 ms per request graph. Random Forest and MLP train in under one second on flattened features, and their prediction latency is lower. Thus, in the current implementation, graph inference is still lightweight enough for offline or near-real-time scoring, but the non-graph baselines are faster and stronger on the present feature set.

For RQ3, logs and metrics both improve over trace structure alone, but the contribution is attack-dependent. Metrics capture performance-visible attacks such as HTTP flood, while logs are expected to be important for semantic probes and improve aggregate performance over trace-only features. For RQ4, graph-based detection is feasible but not yet dominant: the current non-graph baselines outperform the shallow GCN under trial-level split. This does not invalidate graph-based microservice IDS; rather, it shows that graph methods must be evaluated against strong flattened baselines and should incorporate richer edge features, temporal aggregation, and class-aware training before claiming superiority.

\section{Discussion}

The outcome is nuanced. Request-level graph detection is useful for attacks that change call paths, latency, resource context, or security logs. However, attacks that are semantically malicious but operationally quiet may not produce strong performance anomalies. For such cases, log events and richer path-conformance features may provide security-relevant context that metrics alone cannot capture, but the current shallow GCN has not yet demonstrated superiority over strong flattened baselines.

This distinction matters in practice. A request-level graph representation can connect the actual invocation path with performance symptoms and security events, making it a useful substrate for microservice intrusion detection. The present experiments show that this representation is promising, while also showing that graph neural models need richer trace and edge features before they can be claimed to outperform simpler classifiers.

\textbf{Threats to validity.} This study has several limitations. Benchmark microservices may be simpler than production systems. Synthetic attacks may not capture adaptive adversarial behavior. Request labels can be noisy when a request overlaps a scenario boundary or when trace context is missing from logs. The baseline graph-level random split is optimistic because related request graphs from the same trial may appear in both training and test sets, and the extended GCN runs use only 5 epochs for exploratory comparison. Model performance may depend on graph granularity, feature engineering, workload intensity, and telemetry quality. These limitations should be addressed through transparent experiment logs, repeated trials, high-load benign traffic, richer trace and edge features, and sensitivity analysis.

\section{Conclusion}

This paper studies whether graph-based microservice performance detection can be used for microservice intrusion detection. The central position is that performance graph detection is a useful foundation but not a complete replacement for security-aware intrusion detection. It is most effective when attacks disturb service behavior or dependency structure. The current experiments show that logs and metrics improve over trace-only features, but also that strong non-graph baselines outperform the shallow GCN on the present engineered feature set. Future graph detectors should therefore incorporate richer trace attributes, edge features, temporal aggregation, and class-aware training before claiming model-level superiority.

\appendix

\section{Implementation and Reproducibility Details}
\label{app:reproducibility}

The implementation uses Docker Compose deployment manifests, synthetic workload and attack scripts, automated trial collection, run-spec CSV labeling, a baseline GCN classifier, an extended analysis script, and unit tests. The observability stack records metrics, logs, and distributed traces; normalized trial artifacts include trace records, service logs, node metrics, and label windows. Model checkpoints, structured result files, tables, and figures are saved under the experiment output directory. Fault-injection scripts and additional baselines such as Isolation Forest, One-Class SVM, GraphSAGE, GAT, and graph autoencoders remain future work.

Normal traffic is generated by \texttt{experiment/scripts/workload\_normal.py}. Attack traffic is generated by the HTTP flood, brute-force login, SQL injection probe, SSRF-like probe, and exfiltration simulation scripts under \texttt{experiment/scripts/}. Trial collection is orchestrated by \texttt{experiment/scripts/collect\_attack\_trials.py}. The baseline GCN is implemented in \texttt{experiment/modeling/gnn\_classifier.py}; the ablation, trial-level split, non-graph baseline, runtime, t-SNE, and confusion-matrix experiments are implemented in \texttt{experiment/modeling/experiment\_analysis.py}.

The baseline result is reproduced with:

\begin{verbatim}
uv run --with torch --with numpy --with scikit-learn \
  python modeling/gnn_classifier.py \
  --run-spec-file data/labels/bruteforce_existing_run_specs.csv \
  --run-spec-file data/labels/attack_trials_run_specs.csv \
  --epochs 100 --log-dim 16 --hidden-dim 32 \
  --test-size 0.3 --seed 7
\end{verbatim}

The extended experiment artifacts are produced with:

\begin{verbatim}
uv run --with torch --with numpy --with scikit-learn --with matplotlib \
  python modeling/experiment_analysis.py \
  --run-spec-file data/labels/bruteforce_existing_run_specs.csv \
  --run-spec-file data/labels/attack_trials_run_specs.csv \
  --epochs 5 --log-dim 16 --hidden-dim 32 \
  --test-size 0.3 --seed 7
\end{verbatim}

\bibliographystyle{plainnat}
\bibliography{references}

\end{document}